\documentclass[11pt,a4paper]{article}
\usepackage{astron,epsf,epsfig}
\usepackage{doublespace}
\usepackage{graphics}
\usepackage{rotating}

\def\eg{{\it e.g.}\,}
\def\ie{{\it i.e.}\,}
\def\tff{t_{\mathit{ff}}\,}
\def\NHI{N_{HI}\,}
\def\logNHIcm2{\log\left(N_{HI}\,\mathrm{cm^{-2}}\right)\, }
\def\Msun{M_{\odot}\,}
\def\Zsun{Z_{\odot}\,}

\def\figwidth{1.2\textwidth}

\title{The Chemical Evolution of the Universe I: \\
High Column Density Absorbers}
\author{G.P. Mathlin\thanks{Please send offprint requests to A. C. Baker, email: a.baker@astro.cf.ac.uk},
A.C. Baker, D.K. Churches, M.G. Edmunds
\\ Dept. Physics and Astronomy, P. O. Box 913, 
\\Cardiff University, Wales CF24 3YB}

\begin{document}

\begin{spacing}{1.0}

\label{eq:E}

\maketitle
   
\begin{abstract}

We construct a simple, robust model of the chemical evolution of
galaxies from high to low redshift, and apply it to published
observations of damped Lyman-alpha quasar absorption line systems
(DLAs). The elementary model assumes quiescent star formation and
isolated galaxies (no interactions, mergers or gas flows).  We
consider the influence of dust and chemical gradients in the galaxies,
and hence explore the selection effects in quasar surveys.  We fit
individual DLA systems to predict some observable properties of the
absorbing galaxies, and also indicate the expected redshift behaviour
of chemical element ratios involving nucleosynthetic time delays.

Despite its simplicity, our `monolithic collapse' model gives a good
account of the distribution and evolution of the metallicity and
column density of DLAs, and of the evolution of the global star
formation rate and gas density below redshifts $z \sim 3$. However,
from the comparison of DLA observations with our model, it is clear
that star formation rates at higher redshifts ($z > 3$) are
enhanced. Galaxy interactions and mergers, and gas flows very probably
play a major r\^{o}le.

\end{abstract}

\section{Introduction} 
\label{s:intro} 

Observational studies of the formation and evolution of galaxies are
making great strides forward. Unfortunately, understanding of the wide
range of physical processes involved in this key astrophysical problem
is far from complete.  There is a pressing need for a simple yet
robust framework in which to interpret the new observations.

The prevailing view is that galaxies form in dark matter halos that
trace primordial density fluctuations, but subsequent mergers, gas
flows and star formation certainly play an important role in galaxy
evolution.  Great efforts have been made in the `semi-analytic'
approach to modelling galaxy evolution, as summarised by
\cite{BCFL98}. This sophisticated technique uses statistical methods
to follow the growth of dark matter halos, and physically motivated
rules to describe the gaseous and stellar processes within them. {\it
Ab initio} calculations of galaxy evolution in hierarchical
cosmological models are constrained by local galaxy luminosity
functions, enabling statistical and cosmological predictions such as
galaxy survey redshift distributions, and the cosmic star formation
history, to be made.

Our approach is different. We have produced a simple, robust model for
galaxy formation and evolution intended to meet the constraints of the
observed {\it chemical} evolution of the Universe. In particular, in
this paper we investigate the detailed gas phase abundances of quasar
damped Lyman $\alpha$ absorption line systems (DLAs) to $z\sim4$--$5$
(defined as those with neutral hydrogen column densities above
$\logNHIcm2 \sim 20$). We improve upon elementary global calculations
(e.g. \cite{PF95,EP97}) by modelling the enrichment of gas in separate
gravitational potential wells.  We adopt as far as possible the `best'
descriptions of the different physical processes involved, based upon
reliable observations and theory. Our eventual aim is to predict the
outcome of future observations -- particularly in the IR and submm,
with very large telescopes and adaptive optics -- that will allow
chemical analysis of directly detected individual galaxies or galaxy
fragments at high redshifts.

The DLA systems have been linked to proto-galactic clouds
(\cite{Wolfe86}), low surface brightness galaxies (\cite{Jim99}), the
outer regions of galactic disks (\cite{Berg86,Berg91,Stid95}),
sub-galactic star forming regions, and dwarf galaxies
(\cite{Yan92}). DLAs are currently known at redshifts $ 0.1 < z < 4 $,
which probably spans the era of major galaxy formation, and have
observed HI column densities up to $ \logNHIcm2 \sim 21.7 $. We do not
attempt to address the precise nature of the absorbers, but treat them
as clouds of gas (with associated stars and dust) irrespective of what
kind of (proto)-galaxy they may represent.  We have compiled a
database of redshifts, HI column densities and abundances of all the
DLA systems with published measurements.  We have produced synthetic
surveys for overall statistical comparisons, and models of individual
systems, and hence predicted directly observable system properties.

The layout of this paper is as follows. In \S~\ref{s:model}, we
outline our initial model for the formation and evolution of the
galaxy systems, using the Simple chemical evolution model. In
\S~\ref{s:structure}, we discuss the dust optical depth, and add a
simple representation of galaxy internal structure to our model, in
the form of exponential disk gradients in stars, gas, metallicity and
dust.  In \S~\ref{s:structure}, we consider the opacity of galaxies
and possible effects upon the quasar optical surveys.  In
\S~\ref{s:evgaldist}, we use the Holmberg relation for galaxy
luminosity and size to make a preliminary calculation of the evolution
of the galaxy population as a whole. In \S~\ref{s:confrontation}, we
compare detailed model predictions with observations of DLAs,
including the chemical evolution of individual elements, and study
other relevant data such as the properties of Local Group dwarf
galaxies, and the star formation history of the Universe. In
\S~\ref{s:limitations}, we take a critical look at the limitations of
our initial model, and how these might affect our conclusions. In
\S~\ref{s:conc}, we summarise our conclusions.  (Unless otherwise
stated, our calculations assume $H_0=100\,h \, \mathrm{km\,s^{-1} \,
Mpc^{-1}}$ with $h=0.7$, and $\Omega = 1$.)

\section{A Galaxy Chemical Evolution Model}
\label{s:model}

We wish to focus upon {\it chemical} evolution, and so we construct a
simple, yet physically reasonable model of a galaxy, without imposing
particular structural or dynamical properties.  We consider a
spherically symmetric cloud of gas which, at a redshift determined by
its density, breaks away from the Hubble flow and collapses to form a
gravitationally bound system. Given a redshift, the density required
for a region to collapse is (\cite{Peebles93})

\begin{equation}
\label{e:rho0}
\rho_0=\frac{ 0.3 ( z+1 )^3H_0^2\,\Omega_M ( \delta+1 ) }{2G}
\end{equation}

where the density contrast over the mean background density $\delta =
5.6$, and $\Omega_M$ is the mass density of the Universe (for
cosmologies with $\Omega_{\Lambda} = 0$). Guided by detailed 3D
numerical self-gravitating hydrodynamical calculations of the
formation of galactic systems (\eg \cite{ChurT99}), we assume that the
cloud collapses by a factor of 10 from an initial radius $R_0$ over a
free-fall time, halting at $ R = 0.1 R_0$ and evolving at constant
radius thereafter. The free-fall time of the cloud, $\tff$ is

\begin{equation}
\label{e:tff}
\tff= \left( \frac{3\pi}{32G\rho_0} \right)^{\frac{1}{2}}
\end{equation}

where $\rho_0$ is the density of the cloud as it breaks away from the
Hubble flow at the turnaround redshift, $z_{turn}$. The radius of the
cloud, $R$, is given (for $ t < \tff $) by

\begin{equation}
\label{e:r1}
R\sim R_0\left(1-\frac{t}{\tff}\right)^{\frac{1}{2}}.
\end{equation}

We model clouds with total gravitational mass in the range $10^8<
M/\Msun< 10^{12}$. We fix the baryonic fraction to be universally
$10\%,$, so that each cloud contains $90\%$ dark, non-baryonic mass
which does not form stars, nor contribute to the gas column
density. For clouds of uniform density, the initial cloud radius (at
the onset of collapse) is therefore a known function of cloud mass and
the turnaround redshift (Eqn~\ref{e:rho0}).

We use a Schmidt star formation law (\cite{S59}), with the star
formation rate a power-law function of $\rho_{gas}$, the gas volume
density, given by

\begin{equation}
\label{e:gamma}
\Gamma=\kappa\,\rho_{gas}^q
\end{equation}

where $\Gamma$ is the volume star formation density, and $\kappa$ and
$q$ are constants which determine the star formation efficiency.
Kennicutt has shown that such prescriptions are a good representation
of star formation over a wide range of physical conditions in
present-day galaxies (\cite{Kenn88,Kenn98}).  We fix the slope at a
convenient value, $q=1.5$, guided by Kennicutt's prefered slope
$1.4\pm0.1$. We fix the absolute efficiency $\kappa =2.5\,
\mathrm{\Msun^{-1/2} pc^{3/2} Gyr^{-1}}$, which gives
reasonable present-day gas/star mass ratios across a range of initial
model galaxy cloud masses (confirmed by detailed self-consistent
numerical galaxy models (\cite{ChurT99})). The star formation rate in
all our calculations therefore varies only as a function of the gas
density.

The total baryonic mass of the model galaxy cloud is $M_b = 0.1
M$. Initially, just after turnaround, the cloud baryonic mass is
purely gas, so the mass of gas in the cloud $M_g = M_b$, and the mass
of stars, $M_s = 0$.  Star formation proceeds during collapse and
subsequent galaxy evolution and we assume that a fraction, $\alpha$,
of the gas which is consumed by star formation remains in long-lived
stars and stellar remnants. Therefore, the rate of change of the mass
in stars is given by

\begin{equation}
\label{e:dMsdt1}
\frac{\mathrm{d}M_s}{\mathrm{d}t}=\frac{4}{3} \alpha \pi R^{3} \Gamma
\end{equation}

at all times.  Substituting for the volume star formation rate,
$\Gamma$, from Eqn~\ref{e:gamma}, and the current radius of the
cloud, $R$, from Eqn~\ref{e:r1} and noting that $ M_g = M_b - M_s$ we
find

\begin{equation}
\label{e:dMsdt3}
\frac{\mathrm{d}M_s}{\mathrm{d}t}=\frac{\alpha}{V_{0}^{1/2}}\,\kappa\,\left(1-\tff\right)^{-3/4}\left(M_{b}-M_{s}\right)^{3/2}
\end{equation}

for the cloud collapse phase, and

\begin{equation}
\label{e:dMsdt4}
\frac{\mathrm{d}M_s}{\mathrm{d}t}=\frac{1000 \alpha}{V_{0}^{1/2}}\,\kappa\,
\left(M_{b}-M_{s}\right)^{3/2}
\end{equation}

for the constant radius cloud evolution phase, where $V_0$ is the
initial volume of the model galaxy cloud.  Both these equations may be
solved explicitly to give

\begin{equation}
\label{e:mg1}
M_{g}=\left(\frac{2\,\alpha\,\kappa}{V_{0}^{1/2}}\left(\tff-\tff^{3/4}\left(\tff-t\right)^{1/4}\right)+M_{b}^{-1/2}\right)^{-2}
\end{equation}

and

\begin{equation}
\label{e:mg2}
M_{g}=\left(\frac{\alpha\,\kappa}{V_{0}^{1/2}}\left(0.063\,t_{d}+2\,\tff\right
)+M_{b}^{-1/2}\right)^{-2}
\end{equation}

for the collapse and constant radius phases respectively, where $t_d$
is the time since the end of the collapse phase.

\subsection{Chemical Evolution: Testing the Model}
\label{s:modeltests}

The fundamental aim of this work is to make testable predictions about
the chemical evolution of galaxies at high redshift, and in the first
instance, to understand the properties of high column density (damped
Lyman alpha) quasar absorption line systems.  Therefore, we need to
calculate the basic observable properties of galaxies which appear as
DLAs. These are: the absorption redshift $z_{abs}$, the gas phase
metallicity, $Z$ (mass fraction of metals in the interstellar medium)
and the column density in neutral hydrogen, $\NHI$, in the ISM.  The
absorption redshift is simply determined by the turnaround redshift of
the model galaxy cloud, its age, and the cosmology. We can also
readily calculate an average metallicity and column density for any
cloud.

The metallicity can be estimated from the Simple Model of `closed box'
chemical evolution (\cite{P97}).  The standard Simple result is that
the metallicity, $Z$, follows the relation

\begin{equation}
\label{e:z1}
Z = -p \ln f
\end{equation}

where $p$ is the yield (the mass of heavy elements in the processed
gas returning from stars to the ISM per mass locked up in long-lived
stars and stellar remnants), and $f=M_g/M$ is the fraction of baryons
in gas (hereafter, the 'gas fraction'). Throughout our calculations,
we assume a constant, uniform heavy element yield, $p$.

The column density of neutral hydrogen, $\NHI$ can then be estimated
from the cloud gas mass $M_g$ (allowing for primordial helium, and
assuming a typical cloud size $R$) as $\NHI \sim M_g \, R /
(\frac{4}{3} m_H)(\frac{4}{3} \pi \, R^3) $, giving

\begin{equation}
\label{e:colden}
\NHI = \frac{ 9 M_g } { 16 m_H \, \pi \, R^3 }
\end{equation}

where $m_H$ is the mass of a hydrogen atom. This is an upper limit,
since molecular hydrogen will form in the densest regions, and there
will be some ionisation.

We now have a basic framework of galaxy formation and evolution, and
in particular, a robust formalism to study the chemical enrichment
history of any given model galaxy cloud. We next need to study the
dust optical depths that can be expected (\S~\ref{s:structure}) (which
may affect quasar surveys (\S~\ref{s:losing})), and to allow for
internal structure within our galaxy clouds (\S~\ref{s:gradients}).

\section{Gradients and Dust}
\label{s:structure}

A significant fraction of the heavy elements in our Galaxy are locked
up in the solid phase, particularly as small dust grains in the
ISM. We can predict the evolution in the opacity of our model clouds
due to absorption by these grains, $\tau_d$, in terms of $\NHI$ and
gas fraction, $f$. If a cloud is to be detected as a DLA, we must
first detect the background quasar, often in a survey at blue
wavelengths; the dust must not redden and obscure the quasar beyond
survey and spectroscopic magnitude limits (\cite{PF95}).

\subsection{Average Dust Opacity}
\label{s:avedust}

The optical depth of a cloud due to dust $ \tau_d = N_d \, \sigma_d $
where $ \sigma_d \sim \pi r_d^2 $ is the typical grain absorption
cross-section ($r_d$ is a typical grain radius) and $N_d$ is the
column density of dust grains.  For a typical line of sight, the
column density of dust $ N_d \sim n_d \, R $, where $n_d$ is the
number density of grains and $R$ is the characteristic size of the
cloud. Therefore, the dust optical depth can be estimated from the
assumed typical grain properties, and global cloud properties

\begin{equation}
\label{e:taudust1}
\tau_d=n_d\,R\pi\,r_d^2.  
\end{equation}

If $M_d$ is the total mass of dust in the cloud then we can write
(\cite{Edm98}, using our symbols) $ M_d = \eta Z M_g $, where $Z$ is
the mass fraction of metals in the gas phase, $\eta$ is the fraction
of metals which have condensed into dust and $M_g$ is the cloud gas
mass.  We can also write $ M_d = n_d \, m_d \, V $, where $ V =
\frac{4}{3} \pi \, R^3$ is the volume of the cloud; if the typical
density of a dust grain is $\rho_d$ then the typical mass of a dust
grain is $ m_d = \frac{4}{3} \pi \, r_d^3 \, \rho_d$. Therefore,
equating our two expressions for $M_d$,

\begin{equation}
\label{e:dustmass}
\eta Z M_g = n_d\,(\frac{4}{3}\pi)^2\,R^3\,r_d^3\,\rho_d.
\end{equation}

Thus, combining Eqn~\ref{e:dustmass} with the expression for dust
opacity (Eqn~\ref{e:taudust1}), and using the relationship between gas
fraction and metal mass fraction for the Simple Model
(Eqn~\ref{e:z1}), the optical depth due to dust may be written as

\begin{equation}
\label{e:taudust2}
\tau_d= \frac{ 9 M_g \eta p \ln \frac{1}{f} }{ 16 \pi R^2 r_d \, \rho_d }.
\end{equation}

The cloud gas mass and column density are related by
Eqn~\ref{e:colden} so we can eliminate cloud gas mass $M_g$ in the
expression for optical depth (Eqn~\ref{e:taudust2}). The typical dust
optical depth in a model galaxy cloud is therefore

\begin{equation}
\label{e:taudust3}
\tau_d= \frac{ \eta \NHI m_H p \ln \frac{1}{f} }{ r_d\,\rho_d },
\end{equation}

Theoretically, we calculated the gas fraction, $f$, and hence, the
column density, $\NHI$. Observationally, we measure column density and
the metallicity. We can therefore use Eqn~\ref{e:taudust3} to infer
average dust opacities for model galaxy clouds {\it and} observed
systems, if we make some assumptions about the properties of the dust.
Our knowledge of the Galaxy suggests that taking as representative
values $\eta=0.5$ (the fraction of metals in dust), $p\sim0.01$ (the
local stellar yield, $0.5 \Zsun$), $r_d \sim 0.1\mu\mathrm{m}$ (the
radius of a typical `classical' dust grain) and $\rho_d \sim 2000\,
\mathrm{kg\,m^{-3}} \equiv 2 \mathrm{ g\,cm^{-3} }$. With these values,
the typical dust optical depth of a cloud is

\begin{equation}
\label{e:taudust6}
\tau_d \sim 0.4 N_{HI,21} \ln \frac{1}{f}
\end{equation} 

where $\NHI \equiv N_{HI,21} \times 10^{21} \mathrm{ cm^{-2} } $.
This prescription indicates that a model galaxy cloud representative
of our Galaxy, with a gas fraction $f \sim 0.1$ and an HI column
density given by $N_{HI,21}
\sim 1 $ will have a average dust optical depth, $\tau_d \sim 1$. It is
possible that the fraction of heavy elements locked up in dust in the
ISM ($\eta$) is a function of galaxy evolution, but initial estimates
suggest that this is not a significant effect for purposes of the
present calculation (\cite{E01}).

\subsection{Gradients}
\label{s:gradients}

Our calculations so far have implicitly assumed that the dust is
uniformly distributed.  We now allow for radial exponential gas,
metallicity and opacity gradients, as found in disk galaxies.

We assume that the metallicity distribution does not vary azimuthally,
and that $Z( r )$ can be parameterised as $\log Z( r ) =a-br$, where
$a$ is the central metallicity of the disk and the slope, $b \simeq
0.2$ dex/scalelength (\cite{Garn98}).  Equivalently,

\begin{equation}
Z=Z_0\,e^{-0.46r/h_{\ast}}\,,
\end{equation} 

where $h_{\ast}$ is the stellar exponential scalelength of the
disk. We calibrate using the Galactic measurement $Z_{ISM}=0.017$ at
the Solar circle, $r\sim8\,$kpc. Then, if $M_{B,MW}$ is the absolute
blue magnitude of the Galaxy, the central metallicity of a galaxy with
absolute blue magnitude $M_B$, is

\begin{equation}
Z_0=0.027e^{-0.46\left(M_B-M_{B,MW}\right)}.
\end{equation} 

For a typical $\mathrm{L^{\ast}}$ galaxy, the neutral hydrogen is also
distributed in an exponential disk, and the HI scalelength and the
stellar scalelength are roughly the same: $ h \simeq h_{HI} \simeq
h_{\ast} \sim 7\,$ kpc.  Then $N_{HI}(r) = N_{HI_0} \, e^{-r/h}$ and
the opacity gradient is therefore (from Eqn{~\ref{e:taudust3}})

\begin{equation}
\label{e:taudust4}
\tau \left( r \right) = \frac{ 1.3 \times \eta \, m_H }
			{ \rho_d \, r_d } \,
			\NHI \left( r \right)\, Z\left( r \right).
\end{equation}

The central HI column density, and opacity, are uncertain because
conversion to molecular hydrogen may promote the formation of dust,
especially in cooler regions (\cite{VV98}).  Substituting in numerical
values as in the previous section and normalising to the column
density at one scalelength (meaning that $N_{HI_0} = 4 N_{HI_{21}}(h)
\times 10^{21}\, $ cm$^{-2}$) gives

\begin{equation}
\tau\left( r \right) \sim 1.5 \, N_{HI_{21}}(h) \, e^{ -1.46 \, r/h }
\end{equation}

where the column density measure, $N_{HI,h,21}$ is directly calculated
in our model. This relation predicts an average optical depth at the
Solar circle in our Galaxy of $\tau_d \sim 1$, which is again
consistent with azimuthal averages in external galaxies
(\cite{DKRW99}, \cite{KTGW98}).

\section{Line-of-sight Optical Depth}

The analysis of galaxy opacity gradients presented above suggests
that, for a present-day ${L^\ast}$ galaxy, the disk optical depth
becomes low enough to allow the detection of an underlying
${L^\ast_B}$ quasar in typical optical surveys at $\sim 2$ disk
scalelengths; hence $\sim50\%$ of the disk surface area will be
`optically thick' to ${L^\ast_B}$ quasars (see \S~\ref{s:losing}).  We
can therefore set an {\it upper limit} on the fraction of the sky
which is covered by the optically thin regions of disk galaxies.  The
probability $P$ that a random line of sight passes through a given
region of a galaxy varies with redshift as

\begin{equation}
\frac{\mathrm{d}P}{\mathrm{d}z} = n \sigma \frac{ c \left( 1+z \right)^2 }
                                                {  H_0 E( z ) },
\end{equation}

where $ n $ is the number density of galaxies, and $ \sigma $ is the
geometric cross-section of the relevant region of the galaxy
(\cite{Peebles93}). In general, $n$ and $\sigma$ will be (probably
decreasing) functions of redshift, but to determine a firm {\it upper}
limit, we can assume that $n$ and $\sigma$ are constant in
redshift. In an Einstein-de Sitter universe, $ E( z ) \equiv ( 1+z
)^{3/2 }$, and so the probability that a random sightline intersects
the outer regions of a galaxy where there is insufficient dust to
obscure an $ L^{\ast} $ quasar, yet sufficient neutral hydrogen to
give rise to a DLA is

\begin{equation}
P=\frac{3 n \sigma c}{2 H_0} \left( ( 1+z )^{3/2} - 1 \right).
\end{equation}

We take $ n \sim 0.008\, \mathrm{ Mpc^{-3} } $ (\cite{BM98}).  From
our opacity gradient calculations ($\S$~\ref{s:gradients}), the
optically thin cross-sectional area of a typical face-on disk is $
\sim 600\, \mathrm{kpc^2}$. Therefore, our firm {\it upper limit} on
the probability of a line of sight intersecting such a region is $
\sim 10 \% $, for a quasar at an emission redshift $ z_{em} \sim 3
$. Of all the quasars that have been discovered, $ \sim10 \% $ of
quasars which have been spectroscopically surveyed have damped
absorption features (\cite{WLFC95}). This very simple analysis would
therefore suggest that all DLAs are caused by the outer regions of $
L^{\ast} $ galaxies.  How is this conclusion affected by our {\it a
priori} assumptions?  We have assumed that all galaxies are oriented
face-on, which gives an upper bound to the optically thin
cross-section. Implicit in the above reasoning is the assumption that
the population of quasars forms a shell at high redshift (\eg $ z = 4
$), maximising the intersection probability. We assume no evolution in
the number density of the disk population, but if the epoch of disk
formation falls at $ z < 4 $, the number density at earlier times will
be lower. All these factors lead to overestimation of the probability
of intersection. Hence, the covering fraction on the sky of suitable
regions of $ L^{\ast} $ galaxies (optically `thin', `high' $ \NHI $)
is probably {\it insufficient} to explain the frequency of DLAs. The
DLA progenitors are probably a heterogeneous population, with some
absorption features forming in the outer regions of $L^\ast$ galaxies,
but others forming, for example, in the central regions of dwarf
galaxies, or in low surface brightness galaxies.

\subsection{Quasar Selection Biases}
\label{s:losing}

Intervening galaxies with the dust properties discussed above
(\S~\ref{s:structure}) may cause selection effects in quasar and DLA
surveys. These effects will depend upon quasar survey wavelengths and
magnitude limits, and upon spectroscopic follow-up magnitude limits,
and upon the density and luminosity distribution of quasars.

\begin{figure}[p]
\centering
\resizebox{\figwidth}{!}{\rotatebox{-90}{\includegraphics{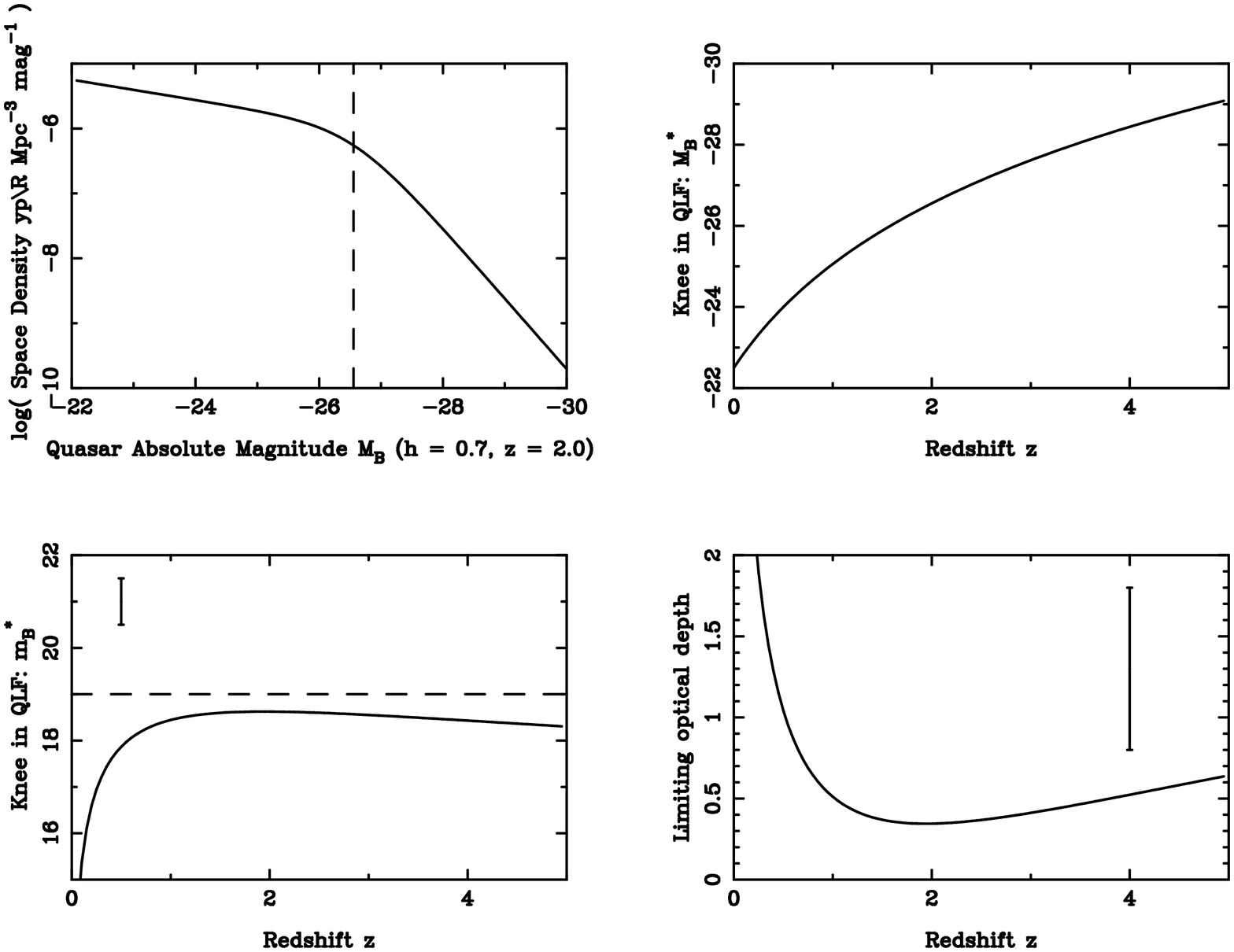}}}
	\caption{\label{f:figure1}}

	% The evolution of quasar optical luminosity function (QOLF), and
	% the effects of obscuration. (a) A parametric fit to the QOLF at
	% redshift $z=2$; (b) A pure luminosity parameterisation of the
	% evolution of the characteristic quasar luminosity (${\mathrm
	% M^*_B}$); (c) The apparent magnitude of the `knee' in the QOLF,
	% accounting for PLE and cosmological dimming (error bar indicates
	% approximate overall uncertainties); (d) The foreground optical
	% depth required to dim an $L_B^*$ quasar below a survey apparent
	% magnitude limit $m_B = 19$ (and uncertainties).}
%
\end{figure}

The form and evolution of the luminosity function of optically
selected quasars is quite well known. The shape of the quasar optical
luminosity function (QOLF) $\Psi \left( L, z\right)$ can be well
parameterised as a double power-law (\cite{BSP88}; see
Fig~\ref{f:figure1}a). The optical QLF has just one characteristic
feature, the `knee' $\Psi^*,{M^{\ast}_B}$, which enables us to
distinguish between luminosity and density evolution in the quasar
population. The QOLF moves monotonically towards higher luminosity as
a power law in redshift, $\propto ( 1 + z )^k$, where the evolution
parameter is $k \sim 3.4$ (see recent determinations by
\cite{FLF97,MP99}, and Fig~\ref{f:figure1}b). Interestingly, this
intrinsic brightening of the characteristic $L^*$ quasar rather
closely balances the effects of cosmological dimming
(Fig~\ref{f:figure1}c).

\vspace{0.25cm}
[INSERT FIGURE 1 HERE]
\vspace{0.25cm}

The QOLF is calculated from quasars discovered in magnitude-limited
blue surveys (\eg the LBQS, \cite{HFC95}). Quasar numbers are
dominated by those fainter than the characteristic luminosity $L^*$ at
all redshifts. The level of foreground obscuration which will dim an
$L^*$ quasar below the typical survey magnitude limit is subject to
uncertainties in the adoption of cosmological and quasar evolution
models. However, for median values of all the parameters, in the
redshift range $ z > 0.8 $ it would take foreground extinction of only
$ \tau
\sim 0.5 $ to push an $L^*$ quasar below an apparent magnitude limit
$m_{B_{\mathrm{lim}}} = 19$ (Fig~\ref{f:figure1}d).

\begin{figure}[p]
\centering
\resizebox{\figwidth}{!}{\rotatebox{-90}{\includegraphics{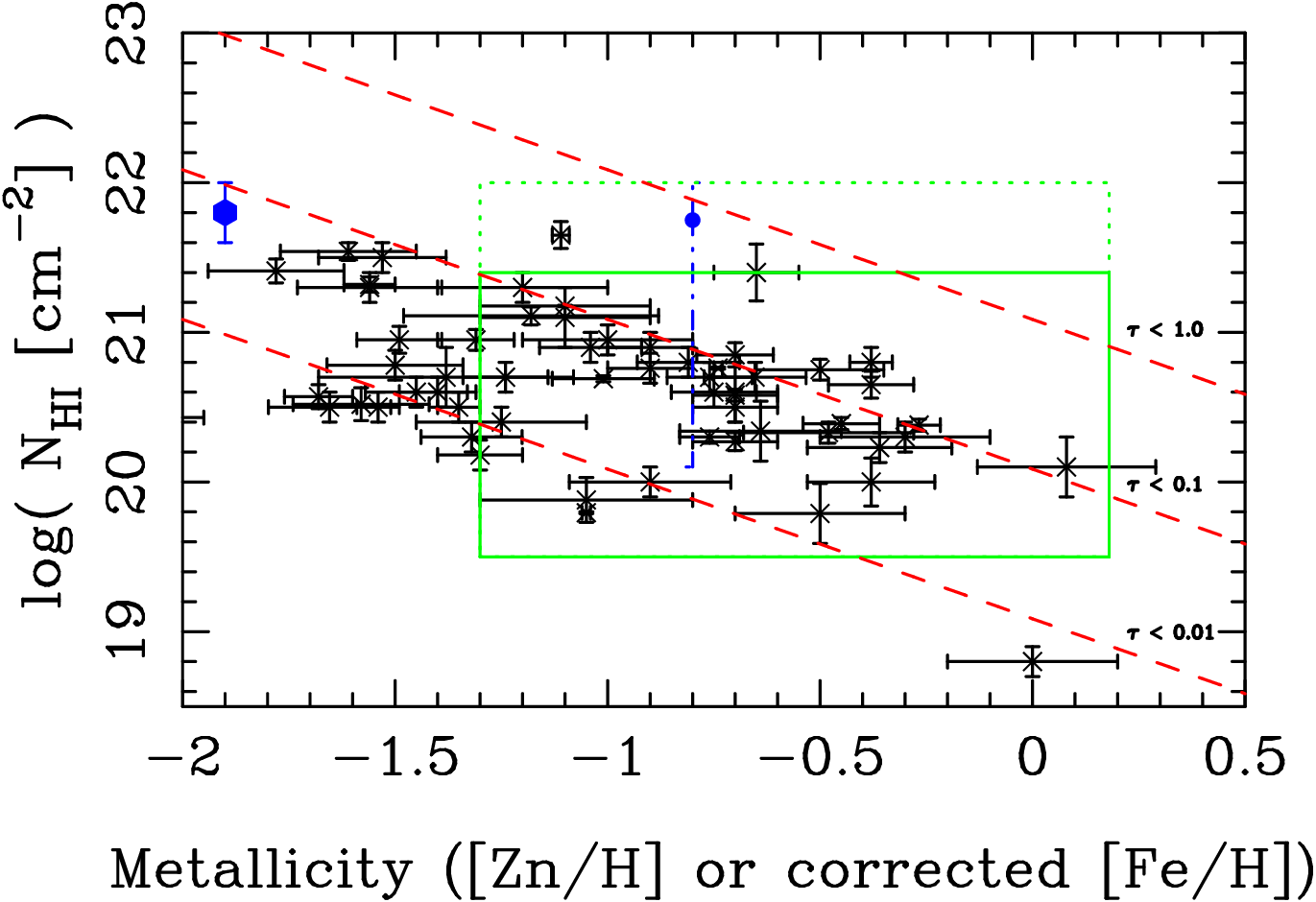}}}
	\caption{\label{f:figure2}}
	% Typical dust opacity predicted from the gas phase metallicity and
	% the neutral gas column density. The dashed lines correspond to
	% typical dust optical depths $\tau_d = 1, 0.1, 0.01$. The solid
	% box show the region occupied by normal galaxies (with a possible
	% extension into the dotted box). The data points at [-1.8, 21.8]
	% and [-1.6, 21.8] are I Zw 18 and the LMC. The remaining data
	% points are all known DLAs for which metallicity measurements are
	% available ([Fe/H] measurements have been corrected for a mean
	% depletion of 0.6 dex (see \S~\ref{s:confrontation}).}
%
\end{figure}

In Fig.~\ref{f:figure2}, we compare the typical opacities
calculated from metallicity and HI column density in
\S~\ref{s:avedust} with observed properties of DLAs and nearby galaxies.
The majority of known DLAs have $ \tau < 0.1$, which will {\it not}
cause a significant number of quasars to be lost from optical quasar
surveys.  However, it is intriguing to note that the DLA data points
cluster along a line of constant opacity, approximated $\tau_d = 0.1$.
Despite other existing evidence to the contrary (\eg Ellison et
al. (in preparation), and see \S~\ref{s:cumulative}, the simplest
interpretation of {\it this} analysis is that many DLAs do exist with
higher opacities, but are missing from our catalogues because they
obscure background quasars beyond optical quasar survey magnitude
limits.

\vspace{0.25cm}
[INSERT FIGURE 2 HERE]
\vspace{0.25cm}

\section{The Evolution of the Galaxy Mass Distribution}
\label{s:evgaldist}

To make statistical statements about the DLA population, we need to
understand the form and evolution of the mass function of galaxies. In
our model, the metallicity evolution of a galaxy is only a function of
its turnaround redshift, which uniquely determines the gas density
(Eqn{~\ref{e:rho0}}).  To break this degeneracy, we need a
relationship between mass (or equivalently, radius), and turnaround
redshift. We can obtain such a relationship (following \cite{PE96}),
from the Holmberg radius-luminosity relation (\cite{H75}):

\begin{equation}
\label{e:Holmberg1}
R_{hb}\left(L\right)\propto L^{s},
\end{equation} 

where $R_{hb}$ is the Holmberg radius, $L$ is the luminosity, and the
parameter $s\simeq 0.4$.  Assuming a constant mass-to-light ratio, the
Holmberg relation implies a mass-radius relation $R_0=A M^s$ ($A$ is a
constant, $R_0$ is the initial cloud radius) and hence a mass-density
relation. We can therefore calculate the (unique) turnaround redshift
for each galaxy mass. Since we are not including galaxy mergers in
this calculation, and we know the present day galaxy luminosity
function, we can construct a history of galaxy formation in the
universe. Large numbers of low mass dwarf galaxies form first, with
higher mass galaxies forming at progressively lower redshifts in ever
decreasing numbers.

We can substitute our mass-radius relation into the expression for the
turnaround redshift (Eqn~\ref{e:rho0}), and re-arrange to obtain mass
as a function of turnaround redshift:

\begin{equation}
\label{e:massred}
M =\left(\frac{\left(z_{turn}+1\right)^3 A^3 H_0^2 
\Omega\left(\delta+1\right)}{2G}\right)^{1/1-3s}
\end{equation}

The overall present-day space density of galaxies of mass $M$ in mass
interval $\mathrm{d}M$ is well modelled by the Schechter luminosity
function

\begin{equation}
\label{e:mfunc}
\Phi( M )\mathrm{d}M =\Phi^{\ast} \left( \frac{ M^{\ast} }{ M } \right)
^{\alpha} \exp\left( - \frac{ M }{ M^{\ast} } \right) \mathrm{d}M
\end{equation}

where $\Phi^{\ast}$ and $M^{\ast}$ are the characteristic galaxy
number density and mass, respectively. The amount of mass per
co-moving unit volume bound up in galaxies which form at a given
redshift is given by

\begin{equation}
\label{e:massev}
\mathcal{M}_{gal}( z_{turn} ) = \Phi^{\ast} M^{\ast} \exp\left( 1 - 
				\left( \frac{ ( z_{turn}+1 )^3 A^3 H_0^2 \Omega ( \delta+1 )}{ 2G } \right)^{1/1-3s}/{M^\ast}\right)
\end{equation}

Unfortunately, this result is extremely sensitive to the slope, $s$,
chosen for the Holmberg relation (Eqn~\ref{e:Holmberg1}). A value of
$s=0.4$ produces a universe populated by galaxies that are too young
and metal poor, whereas $s=0.42$ (Holmberg's own value) produces a
universe populated by over-evolved galaxies. The necessity for fine
tuning of this parameter is unsatisfactory. Another serious problem is
that Holmberg's original tight correlation between radius and
luminosity (\cite{H75}) is much less convincing in large modern
datasets (Gavazzi, private communication). There is considerable
scatter, and selection biases which are caused by the difficulty in
discriminating between stars and high luminosity compact galaxies on
one hand, and insensitivity to low surface brightness galaxies on the
other, suggesting that the Holmberg relation may be no more than a
selection effect. We may be able to side-step these problems entirely
by using the Extended Press-Schechter theory in our galaxy formation
models, and including hierarchical galaxy evolution through mergers
(see \S~\ref{s:limitations}). We will consider this model in our next
paper (Baker et al. 2001). Nevertheless, our simpler ``Holmberg''
model is surprisingly successful in explaining the observed chemical
evolution of DLAs, as we now discuss.

\section{Observations Confront Theory}
\label{s:confrontation}

We are now ready to confront our preliminary theoretical predictions
with the observed properties of damped Ly$\alpha$ absorber systems
from the literature 
%\footnote{References:
%1 Molaro et al. (1998) MNRAS 293 L37
%; 2 Green et al. (1995) ?
%; 3 Pettini et al. (1995) ApJ 451 100
%; 4 Lipman (1995) PhD thesis, Cambridge
%; 5 Lu et al. (1995a) ApJ 447 597
%; 6 Pettini et al.(1998)
%; 7 Lu et al. (1996) ApJS 107 475
%; 8 Junkkinarien (1991) ApJS 77 203
%; 9 Storrie-Lombardi et al. (1996) MNRAS 282 1330
%; 10 Prochaska \& Wolfe (1998) ApJ accepted
%; 11 Prochaska \& Wolfe (1998) ApJ 474 140
%; 12 Centurion et al. (1998) preprint
%; 13 Boisse et al. (1998) A\&A 333 843
%; 14 Pettini et al. (1994) ApJ 426 79
%; 15 Lanzetta et al (1991) ApJS 77 1
%; 16 Matteucci et al. (1997) A\&A 321 45
%; 17 Pettini et al. (1997) ApJ 486 665
%; 18 Meyer et al. (1995) ApJ 451 L13
%; 19 Lanzetta et al. (1995) ApJ 440 435
%; 20 Smette et al. (1995) A\&AS 113 199
%; 21 Meyer \& York (1992) ApJL 399 121
%; 22 Outram, Chaffee \& Carswell (1999) MNRAS 305 685
%; 23 Pettini, Ellison. Steidel, Shapley, Bowen (1999) ApJ
%; 24 Prochaska \& Wolfe (1996) ApJ 470 403
%; 25 Dan Welty's Web Page
%; 26 Aragon-Salamanca, Ellis and O'Brien (1996) MNRAS 281 945
%; 27 York et al. (1991) MNRAS 250 24
%; 28 Petitjean \& Bergeron (1994) A\&A 283 759
%; 29 Lauoresch (1996) PASP 108 641
%; 30 Rao \& Turnshek (2000) ApJS (accepted) astro-ph/9909164v3
%; 31 Centurion et al. (2000) ApJ 536 540
%; 32 Prochaska and Wolfe (2000) astro-ph/0002513
%; 33 Lu et al. (1998) AJ 115 55.}
(\eg absorption line survey data \cite{RT99,SIM96,LWT95,JHB91,LWTLMH91,Wolfe86}, reports of metallicity measurements
\cite{CBMV00,PESSB00,BBBD98,CBMV98,MCV98,MMV97,PSKH97,LSBCV96,SRSRWK95,LSTM95,MLW95,PLH95,PSHK94,MY92},
and metallicity measurements with 8m-class telescopes
\cite{OCC99,PESB99,PW99,PW97,PW96})
%\cite{OCC99,PESB99,PESSB00,PW99,BBBD98,CBMV98,MCV98,MMV97,PSKH97,PW97,
%LSBCV96,PW96,SIM96,SRSRWK95,LSTM95,LWT95,MLW95,PLH95,PSHK94,MY92,JHB91,
%LWTLMH91,Wolfe86}).
%,summarized in Table~\ref{t:catalogue_1}. 
The free parameters of the model clouds are the turnaround or
formation redshift (which fixes the density), and the mass of the
collapsing halo, which together uniquely determine the initial cloud
radius. After commenting upon metallicity indicators
(\S~\ref{s:zinc}), we will now discuss our simulations of DLA survey
data (\S~\ref{s:simsurvs}) in some detail, including
(\S\ref{s:global}) implied global properties (star formation, gas
density). In \S~\ref{s:fits}, we examine fits to individual DLAs, then
(\S~\ref{s:LGdwarves}) briefly compare the properties of DLAs with
dwarf galaxies in the Local Group. In \S~\ref{s:delays} then consider
the chemical evolution of time-delayed species.

\subsection{Gas Phase Metallicity Indicators}
\label{s:zinc}

Zinc has been considered as a fair and useful indicator of the gas
phase metallicity because Zn is measured to follow Fe in Galactic
stars, suggesting that Zn might track the stellar nucleosynthesis of
Fe in Type Ia supernovae (\cite{AT85,TA85}).  However, recent
measurements in metal poor and thick disk stars find a ratio of
[Zn/Fe] $\sim 0.1$ (\cite{PNCMW00}), perhaps consistent with an origin
in Type II supernovae (\cite{A96}). Whatever its stellar origin, Zn is
expected to be far less depleted onto dust grains than Fe
(\cite{Pet99}). Unfortunately, the weakness of the observationally
accessible ZnII transitions, whilst circumventing problems of
saturation, means that Zn is extremely difficult to measure at
$z_{abs}>3$ (\cite{PW00}). Current measurements of DLA abundances show
$<$[Zn/Fe]$> \sim 0.6$, whereas $<$[Zn/Cr]$> \sim 0.3$, implying
depletion factors of 4 and 2 for Fe and Cr respectively {\it if} Zn is
undepleted. As accurate measurements of [S/Si] become available for
DLAs with [Zn/Cr] $< 0.3$, it will be possible to test for
metallicity-dependent depletions (\cite{PESSB00}). In the meantime, we
too will use Zn as our primary metallicity indicator.

%\include{dla_data_table}
%\addtocounter{table}{1}

\subsection{Simulated DLA Surveys}
\label{s:simsurvs}

\begin{figure}[p]
\centering
\resizebox{\figwidth}{!}{\rotatebox{-90}{\includegraphics{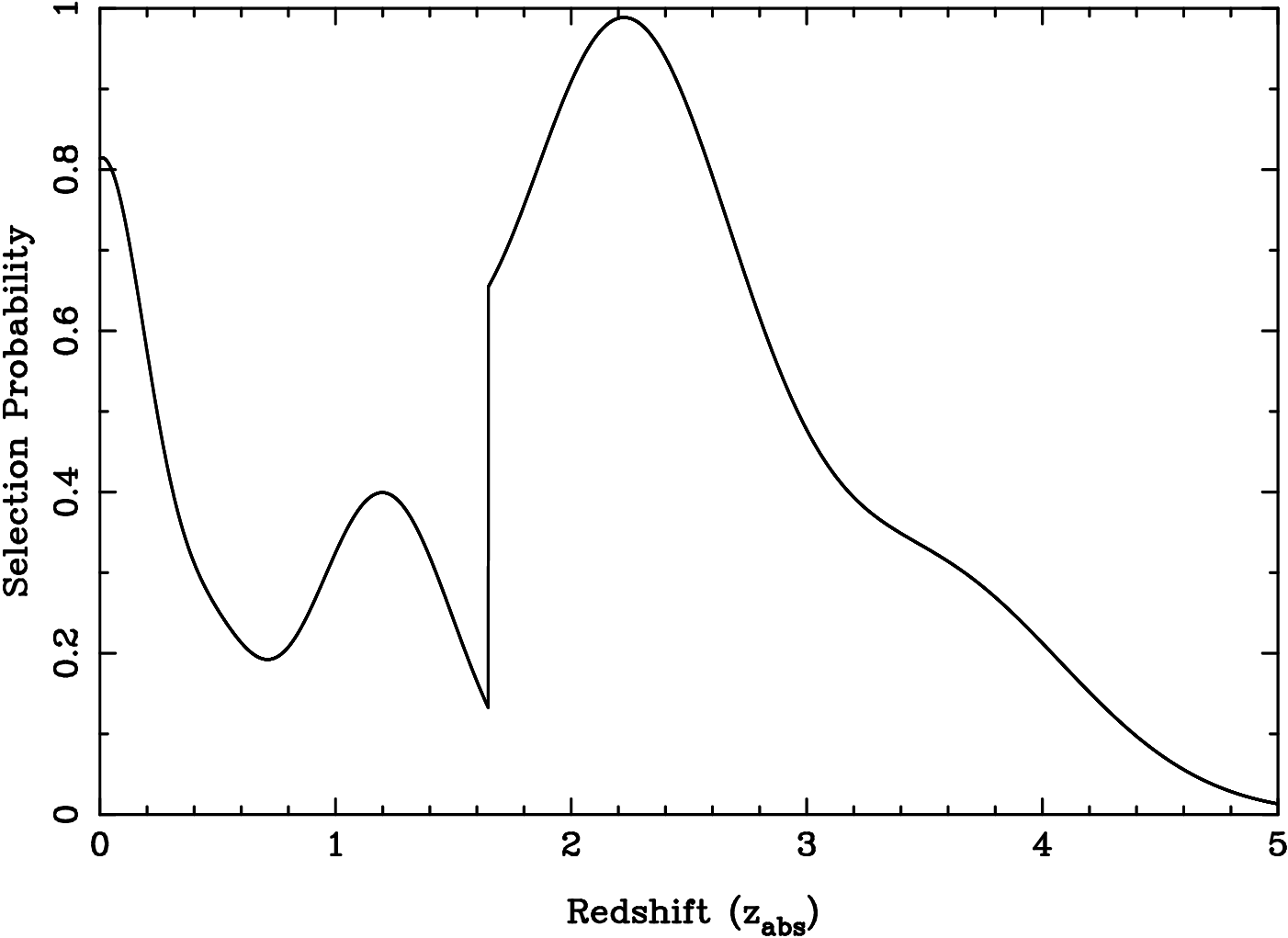} }}
    \caption{\label{f:figure3}}
	% The model redshift selection function, which is a fit to the
	% combined redshift survey path lengths, X(z), of the existing
	% large, well defined DLA surveys (see text).}
%
\end{figure}

\begin{figure}[p]
\centering
\resizebox{\figwidth}{!}{\rotatebox{0}{\includegraphics{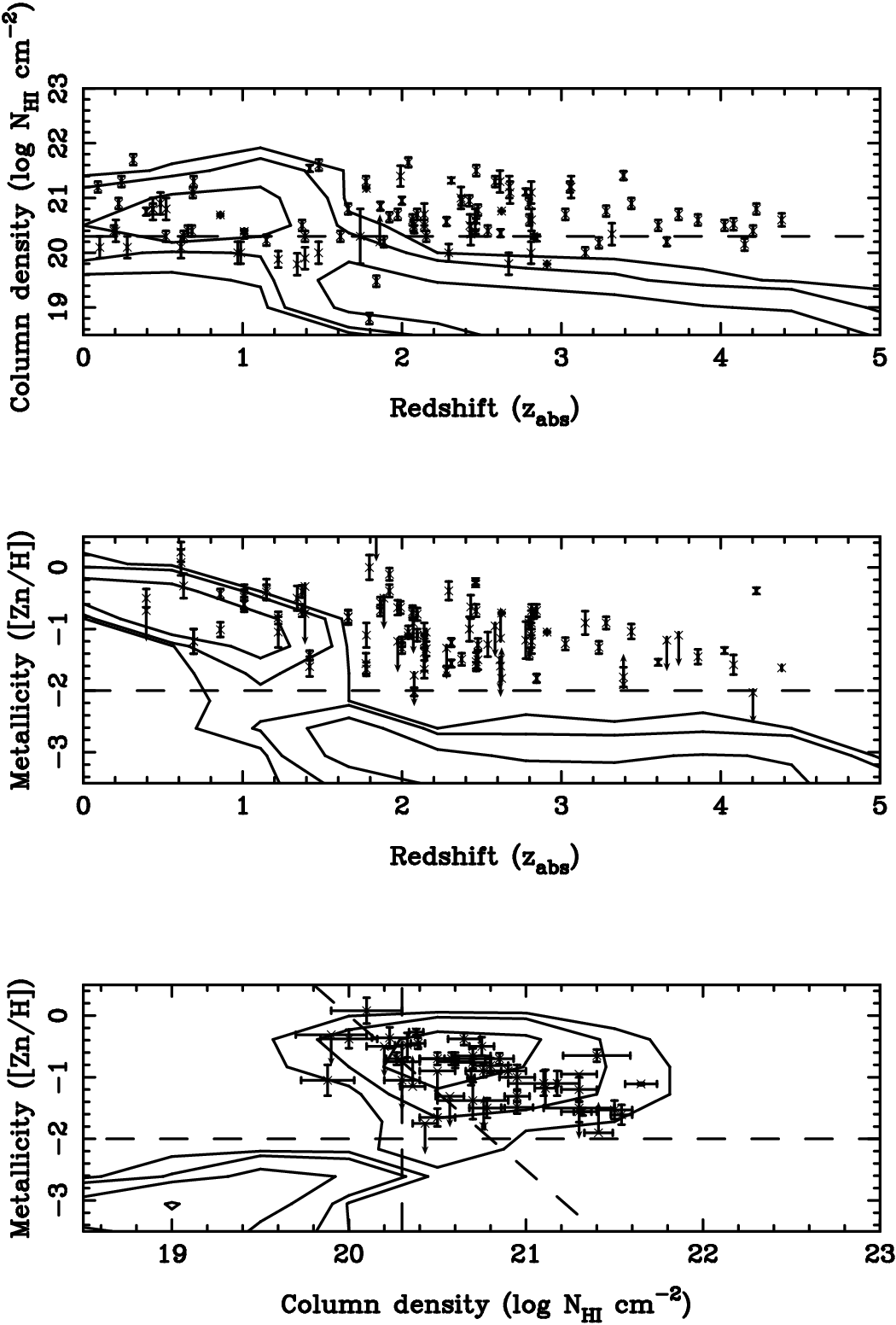} }}
    \caption{\label{f:figure4}}
 	% Comparison of DLA observations (points) with the predictions of
	% our model (contours) using Monte Carlo simulations in Einstein-de
	% Sitter cosmology. The dashed lines indicate the approximate
	% selection effects which affect the data (derived from DLA
	% surveys, including the metallicity measurements of Pettini and
	% collaborators; see text).}
%
\end{figure}

The principle observations which we are attempting to explain are the
redshifts, column densities and metallicities of DLA systems.  To
compare our model with these observations, we created artificial
surveys of 1000 DLAs, including overall observational selection
effects in redshift, column density and metallicity. We consider our
model in two simple yet representative cosmologies: the Einstein-de
Sitter (EdS; $\Omega_M = 1$), and a low density cosmology with
$\Omega_M = 0.3$, $\Omega_k = 0.7$; $\Omega_{\Lambda} = 0$ and $h =
0.7$ in both cases\footnote{We assume that the contributions of
matter, cosmological constant and curvature are such that $\Omega_M +
\Omega_{\Lambda} + \Omega_k = \Omega_{total} = 1$.}. We have sampled
the model DLA galaxies at random impact parameters along the gas and
metallicity radial gradients. The results, in terms of $z_{abs},
N_{HI}$ and [Z/H], are summarised in Figures~\ref{f:figure4} and
\ref{f:figure6}, with the observed data points, and simulated survey
number density contours.  Absorption redshift is the random input
parameter for each model DLA system, and we use the redshift selection
function shown in Fig.~\ref{f:figure3}, representing the combined
redshift survey path lengths $X(z)$ of the DLA surveys of
\cite{LWT95,SMI96} and \cite{RT99}).

It is clear that, in the EdS cosmology, our model is a poor
description of the data for redshifts $z>1.8$ (see
Fig.~\ref{f:figure4}), and we will concentrate on our other
representative model, the `low density' cosmology, in the rest of this
discussion.

\vspace{0.25cm}
[INSERT FIGURE 3 HERE]
\vspace{0.25cm}
[INSERT FIGURE 4 HERE]
\vspace{0.25cm}

For HI column density, the observational detection threshold is set at
$\logNHIcm2 > 20.3$ in all statistically complete DLA surveys. For
metallicity, the sensitivity limit depends upon HI column density, and
there is no uniform Zn equivalent width limit applied to observational
surveys. Therefore, we have estimated the effective metallicity
selection function from the surveys of Pettini et al. to be
[Zn/H]$_{min} \sim -2.5 \logNHIcm2 + 50$ (\cite{PESSB00} and
references therein; \cite{PW00} quote a similar combined limit, that
$\logNHIcm2 + $ [Zn/H] $ > 19.0$). These selection functions are
indicated in Fig.~\ref{f:figure4} as dashed lines, and have been
folded into Figs.~\ref{f:figure5} and
\ref{f:figure6}.

\subsubsection{Cumulative distributions}
\label{s:cumulative}

\begin{figure}[p]
\centering
\resizebox{\figwidth}{!}{\rotatebox{0}{\includegraphics{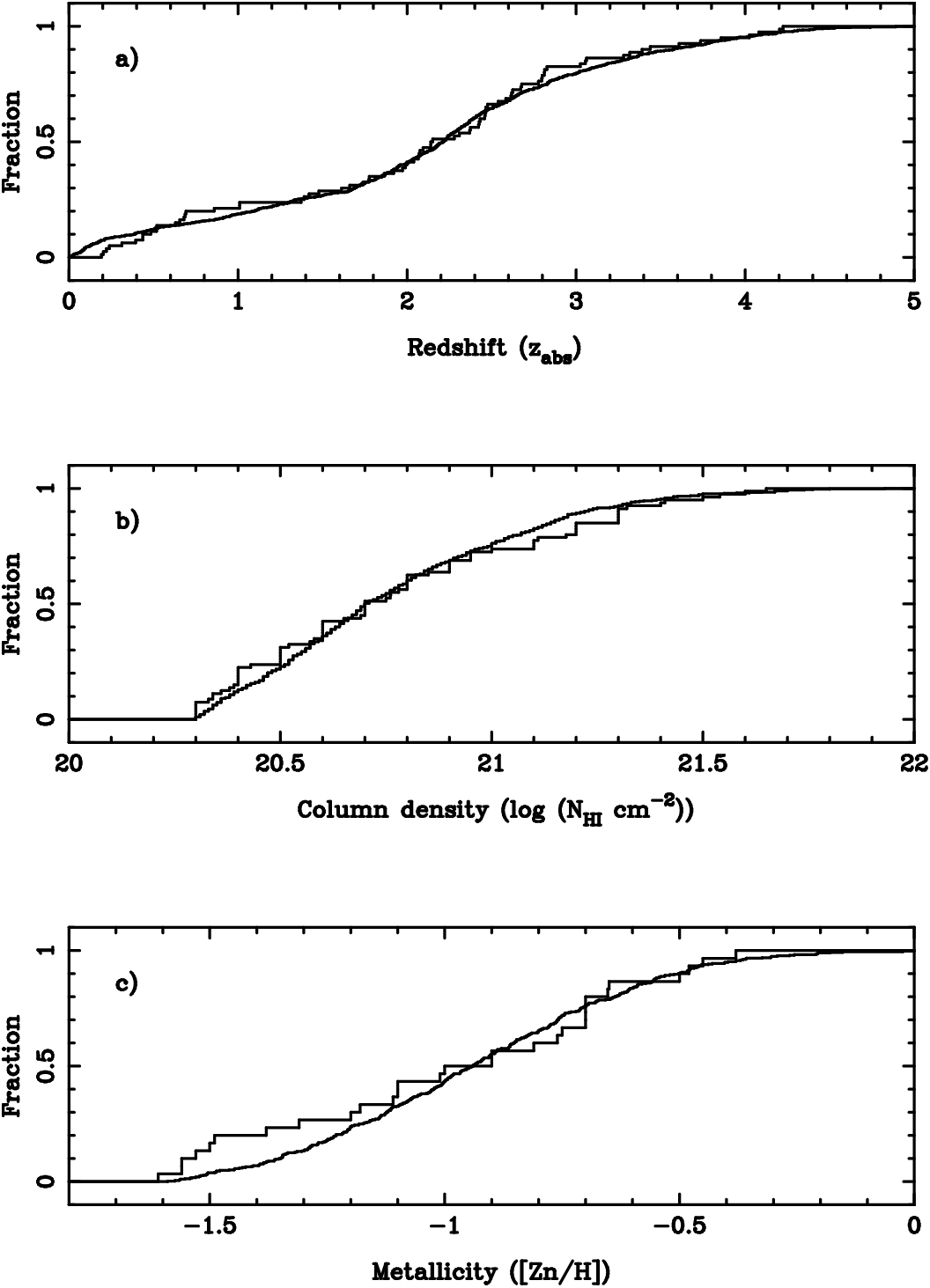} }}
    \caption{\label{f:figure5}} 
	% Cumulative distributions of observed (n=81) and synthetic (n=391)
	% DLAs in (a) redshift, (b) column density and (c) metallicity in
	% the low density universe. Identical selection effects,
	% empirically derived from DLA surveys (including the metallicity
	% measurements of Pettini and collaborators; see text), have been
	% applied to the model and the data. }
%    
\end{figure}

The cumulative distributions of redshift, column density and
metallicity for the low density cosmological model are shown in
Fig.~\ref{f:figure5}.  The smoother line is the distribution of
simulated DLA systems and both curves are normalised to show the {\it
fraction} of `surveyed' systems included up to a given parameter
value. The observed and model cumulative redshift distributions agree
well.

The column density measurements are subject to a Malmquist type bias,
but for the column density distribution function calculated by
\cite{SIM96} and the existing absorption redshift survey path lengths,
$X(z)$, only about one spurious extra DLA system is expected to be
scattered into the lowest column density bin. Again, the model and
observed column density distributions agree well.  

The cumulative metallicity density distributions
(Fig.~\ref{f:figure5}c)), show that we have a deficit of low
metallicity systems ([Zn/H] $< -1.3$). This may relate to star
formation at high redshifts, a point which we return to in
\S~\ref{s:global}.

The predictions in the low density cosmology are clearly a
significantly better representation of the data that those in the EdS
cosmology, especially at redshifts $z < 3$ where the Zn measurements
exist. The well-known observed deficit in high HI column density
systems ($\logNHIcm2 > 21.0$) at high redshifts ($z > 3.5$) is
reproduced (Fig.~\ref{f:figure6}a), because the sequence of
turnaround redshift in our model is a strict function of galaxy
mass. The metallicities are well described by this model, including
the lack of strong correlation of metallicity with redshift at the
current level of observational accuracy (\cite{PSKH97}; our
Fig.~\ref{f:figure6}b). Despite our deficit of low metallicity
systems, we predict that there are numerous absorbers with
metallicities in the range $-2.0 < $ [Zn/H] $< -0.8$ waiting to be
discovered with column densities $19.5 < \logNHIcm2 < 20.5$.  We also
find that {\it all} our model absorbers have internal dust opacity $
\tau < 0.5 $ (with $ 50\%$ below $\tau = 0.1$). This result can be
interpreted in two very different ways. Taking our current model at
face value, we predict that there are no significant selection effects
in optical quasar surveys due to obscuration in foreground galaxies
(see
\S~\ref{s:losing}). However, the current model does not consider
interacting, starbursting systems which are expected to have much
higher opacity, and which we will examine in a future paper
(\cite{B01}).

\vspace{0.25cm}
[INSERT FIGURE 5 HERE]
\vspace{0.25cm}
[INSERT FIGURE 6 HERE]
\vspace{0.25cm}

\subsubsection{Global Properties of the Universe}
\label{s:global}

The evolution of the global properties of the Universe (as previously
calculated \eg by \cite{WLFC95,SMI96,SAGDP99}) emerge `for free' from
our model. The star formation history is shown in
Fig.~\ref{f:figure7}, with data points derived from UV observations
(as summarised by \cite{SAGDP99}, {\it without} attempted dust
corrections) shown for comparison with our independently derived
smooth prediction.  Our simple model is a remarkably good
representation of all the data at redshifts $ z < 4$, but the real
Universe appears to form more stars more quickly at the earliest
observable times. Therefore, we would expect our high redshift ($z>4$)
metallicities to underpredict the observations, which indeed they do,
as shown in Fig.~\ref{f:extra_figure}.

\vspace{0.25cm}
[INSERT FIGURE 7 HERE]
\vspace{0.25cm}

Our detailed prediction of the star formation history depends upon the
adopted Holmberg relation and star formation parameter values. The
observations are at optical and UV wavelengths, which are particularly
sensitive to steady, normal star formation or star formation bursts
with low dust obscuration, and are strongly biased against heavily
dust-embedded star formation, as seen in interacting and starbursting
systems. Such embedded stellar energy generation is certainly
comparable to that detectable at UV/optical wavelengths
(\cite{Puget96,Hauser98}, and may be particularly important at higher
redshifts ($z>1$). Our models do not include any effects of gas
inflows, such as those triggered by interactions and mergers of
galaxies, which appear to be the driving force behind dust embedded
starburst activity. Therefore, it is particularly interesting that the
global star formation history of our model (with steady, non-bursting
star formation) is broadly similar, for redshifts $z<3$, to the best
UV/optical observations currently available. We compare our
calculations to the UV/optical data {\it without} the uncertain
observational dust reddening corrections, both because these
corrections are so uncertain, and because they are likely to be small
in UV-bright star formation regions. Galaxy evolution calculations
which account for interactions, mergers and gas flows are required to
study to {\it true, total} star formation history of the Universe.

\vspace{0.25cm}
[INSERT FIGURE 8 HERE]
\vspace{0.25cm}

In Fig.~\ref{f:figure8}, we summarise the evolution of the baryons in
our model clouds, as distributed between gas, stars, metals and dust
(Fig.~\ref{f:figure8}). Our prediction of the total gas contribution
in bound systems (dashed line) is in reasonable accord with the data
(points from \cite{SMI96}), which are based upon observations of HI in
DLAs. Again, the data are consistent with more star formation at
$z>4$.

\vspace{0.25cm}
[INSERT FIGURE 9 HERE]
\vspace{0.25cm}

\subsection{Detailed fits to individual DLAs}
\label{s:fits}

We now move from broad-brush, statistical comparisons
($\S$\ref{s:simsurvs}) to detailed study of individual DLA features.
We can only model systems in which the absorption redshift, $z_{abs}$,
the HI column density, $N_{HI}$ {\it and} the metallicity ([Z/H], as
measured by [Zn/H], or [Fe/H] corrected for $<$[Zn/Fe]$>_{DLA}$) are
known.  We then take a slice in redshift space though our simulated
universe at $z_{abs}$ (the most precisely determined observable) and
extract all the galaxies with $N_{HI}$ and [Z/H] which are consistent
with the observations. This gives sets of model galaxy parameters,
including turnaround redshifts, masses, and impact parameters to the
QSO line-of-sight. The results of our galaxy chemical evolution models
for each observed DLA yield detailed star formation histories, which
can then be used with a stellar spectral synthesis code (\cite{BC96})
to predict observable galaxy properties at the DLA absorption
redshift, and indeed, any other epoch.

We can now select a well-defined sample of DLAs for further
observational followup.  Of the 63 DLAs with complete ($z_{abs}$,
$N_{HI}$, [Z/H]) data, we could simultaneously obtain good fits to the
observed redshift, column density and metallicity of the DLA galaxy in
18 ($\sim 30 \pm 10\%$) cases (\ie within the quoted (1 $\sigma$)
errors). It is likely that many of the other (unfitted) DLAs have
undergone significant bulk gas flows, possibly involving merger-driven
star formation events.  For those DLAs which we can fit well, the
galaxy masses are $ M \sim 3 \times 10^9$ -- $ 3 \times 10^{10}
M_\odot$, the galaxies are at impact parameters of $b < 2\, $kpc, and
will have apparent magnitudes of $17 < K < 28$.  We therefore predict
that a large fraction of DLAs are caused by sub-$L^{\ast}$ galaxies
sitting almost directly in front of the QSO. This prediction implies
that the $\sim L^\ast$ galaxy associated with the absorber
Q2233$+$1310 ($z_abs = 3.151$) by Steidel and others
(\cite{SPH95,DPBE96}), about $15 h^{-1}$ kpc from the quasar sight
line (in the same cosmology as our individual model fits) is either
atypical, or is not the true absorber.

For each DLA we can fit well, we have selected up to three candidate
model galaxies which bracket the plausible mass distribution, and
evolved these from formation to their absorption redshift, using
stellar synthesis to estimate key observables such as the galaxy
K-band apparent magnitude (in the appropriate cosmology).  The
predicted galaxy properties for each DLA are summarised in
Table~\ref{t:results_0}, with typical uncertainties being $\pm
0.5\,$kpc in the impact parameter and $\pm 0.5$ in the absolute and
apparent magnitudes.

\vspace{0.25cm}
[INSERT TABLE 1 HERE]
\vspace{0.25cm}

As would be expected from our galaxy formation prescription, galaxy
luminosity is broadly anticorrelated with (absorption) redshift. We
predict that some DLA galaxies will be too faint and too close to the
quasar to be detectable even with near-IR detectors on 8m
telescopes. But there is a small, accessible population with predicted
$K < 22$, and $b > 2\,$kpc at redshifts $z < 2.5$.  The
optical--infrared colours are generally moderate (B$-$K $<$ 5.5), but
up to a third of these DLAs arise in rather red (6 $<$ B$-$K $<$ 8.5)
objects, typically at redshifts $ 2 < z_{abs} < 3 $.

These results demonstrate the importance of detailed galaxy chemical
evolution modelling in the design of large telescope programmes to
observe DLAs and galaxies. For example, \cite{AEO96} (AEO) use a
4m-class telescope to image 10 DLA quasars at K, and found two {\it
candidate} DLA galaxies, towards Q$0841+129$ (which has 3 DLA systems)
and Q$1215+333$, with $ K \sim 20 $ and $ b \sim 10\, $kpc; they set
magnitude limits of $ K
\sim 20.3 $--$20.7$ for the others. We predict that the true absorbing
galaxy towards Q$1215+333$ ($z_{abs} = 1.99$) has a mass $M_b \sim 5
\times 10^{9} \Msun$, with $K \sim 22 $ and impact parameter of only $b \sim
1\,$kpc. The closest DLA towards Q$0841+129$ is at $z_{abs} = 2.37$,
and we predict an absorbing galaxy with mass $M_b \sim 5 \times 10^{9} \Msun$, with $K \sim 21.5$ and impact parameter of only
$b \sim 1\,$kpc (the $z_{abs}=2.48$ system will be very similar).  The
other DLAs in the AEO sample are also probably sub-$L^{\ast}$ galaxies
at small impact parameters, well beyond the apparent magnitude and
angular resolution limits of AEO. We therefore do not believe that AEO
have (nor could have) detected the true DLA absorbers in the DLAs they
studied. Without spectroscopic galaxy redshifts, we can only speculate
that the AEO galaxies may be in the same cluster as the true DLA
absorbing galaxies.

\subsection{Local Group dwarf galaxies}
\label{s:LGdwarves}

We can do `backyard cosmology' by comparing the properties of the
dwarf galaxies in the Local Group with our models.  If we believe that
dwarf galaxies give rise to a significant fraction of the observed
DLAs, then we must explain why many dwarfs in the Local Group are too
gas-poor to produce the column densities in DLAs. This may be an
environmental effect.  The Local Group dwarf galaxies clearly separate
into two classes (\cite{Mateo98}): gas-poor dwarfs, those within $\sim
250\,$kpc of the Milky Way or Andromeda; and gas-rich dwarfs, which
lie further out, as shown in Fig~\ref{f:figure9}a. Dwarfs near giants
have probably had their gas stripped by tidal interactions.

The stellar metallicities in the Local Group dwarf galaxies are the
mean values measured for old and intermediate-age stars
(Fig~\ref{f:figure9}b), and thus reflect the gas-phase metallicities 1
-- 10 Gyr ago. The observed values, $ -2.0 < $ [Fe/H] $ < -0.5 $, are
similar to the properties of DLAs at redshifts $ z \sim 2 $, which in
our low density cosmology corresponds to a look-back time $ \sim 6 $
Gyr. There may be a hint that the lowest mass dwarf galaxies ($< 10^8
\Msun$) have the lowest metallicities, although the circum-Galactic
dwarfs dominate the statistics. Low mass galaxies may be subject to
the suppression of star formation and hence chemical evolution due to
supernova winds, leaving gas-poor, low metallicity
systems. Alternatively, low mass galaxies may simply have very low
star formation rates, giving a gas-rich, low metallicity system. The
gas fraction of the majority of Local Group dwarf galaxies are
comparable to those inferred for DLA absorbers from our models.

\vspace{0.25cm}
[INSERT FIGURE 10 HERE]
\vspace{0.25cm}

We therefore conclude that there {\it are} dwarf galaxies in the bulk
of the volume of the Local Group which have gas fractions and
metallicities compatible with those of DLAs. Further investigation
would be desirable, but we do not attempt the necessary more detailed
modelling of Local Group dwarfs here.

\subsection{Time delayed elements}
\label{s:delays}

The main sources of heavy elements are thought to be Type II
supernovae, which enrich the interstellar medium `promptly' on
timescales of $\sim 5 \times 10^6$ yr, and Type Ia supernovae, which
are `delayed' by up to $\sim 1$ Gyr. The elements predominantly
produced in massive stars are thought to include Si, O, Mn and Zn. It
is thought that Fe is partly produced in SNII, but that SNIa may be at
least twice as efficient Fe producers. We have therefore modelled the
evolution of element {\it ratios} by calculating a delayed metallicity
as well as the standard prompt metallicity which we compare with
[Zn/H] measurements throughout this paper. The evolution of
`Promptium' (which can masquerade as \eg\, O, S, Si) with respect to
an element, `Delaydium', which is one third prompt and two thirds
delayed (\eg Fe) can be calculated for different relative time delays,
such as 0.25, 1 and 3 Gyr.

\vspace{0.25cm}
[INSERT FIGURE 11 HERE]
\vspace{0.25cm}

At early times (high redshift, low metallicity) in the model systems,
only the prompt portion of iron has been produced, and so [Pr/De]
$=\log \left( \frac{1}{3} \right) = 0.47$. After $\tau = 1$ Gyr, the
delayed portion of iron starts to build up and [Pr/De] falls towards
the final cosmic value, [Pr/De]$=0$. To reach higher values such as
[Pr/De]$\sim 1$ would require that Delaydium only has \eg $\sim 10\%$
prompt component.

In Fig.~\ref{f:figure10}, we compare the model predictions for the
build-up of iron (Delaydium) relative to oxygen, sulphur and silicon
(Promptium) for a time delay $\tau = 1$ Gyr in the low density
cosmology with the best currently available data, as a function of (a)
absorption redshift and (b) metallicity as measured by iron
(Delaydium). These data are clearly subject to the usual uncertainties
about depletions, and the putative prompt production
pathways. Ironically, the most theoretically useful prompt element is
oxygen, for which the observations are still subject to particularly
significant ambiguities (\cite{PW00}). At this stage, all we can say
is that our synthetic DLA survey predictions are in reasonable
agreement with the available data. More informative investigations of
element ratios must await abundance measurements with an accuracy
$\sigma_{\mathrm{ [Z/H] }} \sim 0.1$, which are starting to become
available (\cite{PW00}).

\section{Limitations of the Model}
\label{s:limitations}

We have made no attempt in this paper to model the r\^{o}le of mergers
in the formation and evolution of galaxies.  However, it is clear from
infrared surveys (which are efficient at finding interacting galaxies
(\cite{CSMS96})), and from theoretical work (\cite{BH96}), that
mergers are a fundamental galaxy formation process.  Why is it
therefore that our model galaxies do so well in reproducing the
observational data for DLAs? As a step towards answering this
question, we outline a scenario in which mergers {\it do} play a
r\^{o}le.

Suppose that the first generation of bound objects in the Universe
were small sub-galactic units of mass $\sim 10^6$ -- $10^7 \Msun$
which somehow produce stars. These low-mass star forming units are
likely to lose their gas in supernova-driven winds after only a few
generations of star formation.  If we allow some of these gas-poor,
low-mass stellar systems to merge, they would form larger bound
systems, embedded in halos of lightly processed ejected gas. A
fraction of this lightly enriched halo gas could settle into a disk
around the merged stellar systems, and more familiar population I and
II star formation could proceed.

The statistics of the merging systems and the subsequent synthesis of
giant galaxies can be described by the Press-Schechter model of
galactic structure formation (\cite{PS74}). The interesting point
about the scenario outlined above, is that only a small amount of star
formation takes place prior to the formation of the disk galaxy.  In
our simple model, very little star formation takes place during the
collapse phase. This suggests that the present day state of a giant
galaxy as predicted by this type of model may be insensitive to the
details of its formation mechanism (\cite{KG91}), and hence, our
preliminary model can still appropriately describe the data.

\section{Conclusions and Next Steps}
\label{s:conc}

We have constructed a simple, robust model of galaxy chemical
evolution, which we quite successfully applied to the study of
galaxies detected as damped Ly$\alpha$ systems in the spectra of
quasars. We have modelled the evolution of neutral hydrogen column
density, metallicity and hence, dust opacity, in clouds of initially
primordial gas from the point where they begin to collapse under self
gravity, up to the present.  We find that the outer regions of model
giant galaxies, {\it and} model sub-$L^\ast$ galaxies are able to
produce absorbers with column density and metallicity typical of DLAs,
and remain optically thin. We have studied the specific properties of
individual DLAs and find that the models of the underlying galaxies
are in good agreement with the limited imaging observations which
currently exist, making sense of the relative lack of success in
detecting DLAs (\eg \cite{AEO96}). Contrary to some expectations
(\cite{WLFC95}), we predict that a significant proportion of the DLA
systems at moderate redshifts ($z\sim2.5$) arise in sub-$L^\ast$
galaxies.  We have therefore initiated a new programme of infrared
imaging using the Gemini North telescope, which is capable of
detecting these fainter galaxies at smaller impact parameters around
DLA quasars.

Our `monolithic' models provide a reasonable account of the chemical
evolution of the Universe at redshifts below $z\sim3$. However, in
attempting to use the Holmberg relation to constrain the cosmic
history of galaxy formation, we have exposed the limitations of our
model assumption concerning the unimportance of galaxy mergers,
particularly in promoting star formation at high redshifts ($z > 3
$). Our next step is therefore to a slightly more sophisticated galaxy
formation scenario, including mergers, to search for observable
effects, including changes in predicted DLA properties.

\vspace{0.25cm}
\begin{flushleft}{\Large\bf Acknowledgements}\end{flushleft}
\vspace{0.25cm}

The authors thank Max Pettini for stimulating and useful discussions,
Dave Clements for comments on draft versions of this MS, and Roger
Philp for advice in code optimisation. ACB is supported by PPARC and
GPM acknowledges a Cardiff University Physics and Astronomy research
studentship.

\bibliography{refs}
\bibliographystyle{astron}

\pagebreak

\begin{table}[p]
\caption{\label{t:results_0} Predicted properties of the absorbing galaxy from successful fits to individual DLAs. (Assumed cosmology: \hbox{$\Omega_M=0.3$}, \hbox{$\Omega_k=0.7$}, \hbox{$\Omega_\Lambda=0.0$} (`low density')).}
\begin{tabular}{ccccccccccc}
         DLA & Absorption  & Impact    & Mass      & Mass                        &  &   &  \\ 
        name & redshift    & parameter & (baryons) & (stars)                     &  &   &  \\ 
             & $z_{abs}$		&$b$ ($h^{-1}$kpc)&  $\log(\Msun)$ & $\log(\Msun)$ & $M_{bol}$ &   $m_K$      &   $B-K$     &    \\ 
0100$+$13000 &  2.309		&   $ 0.8$   &    $10.5$   &    $ 8.0$   &    $-22.2$   &    $19.4$   &    $ 5.4$   \\ 
0112$+$03000 &  2.423		&   $ 0.0$   &    $ 7.9$   &    $ 6.0$   &    $-15.3$   &    $27.6$   &    $ 8.5$   \\ 
0149$+$33500 &  1.613		&   $ 1.4$   &    $10.0$   &    $ 7.8$   &    $-20.6$   &    $21.1$   &    $ 5.5$   \\ 
0454$+$03600 &  0.860		&   $ 0.9$   &    $10.0$   &    $ 8.3$   &    $-19.0$   &    $22.3$   &    $ 5.5$   \\ 
0528$-$2500A &  2.140		&   $ 0.8$   &    $ 9.7$   &    $ 7.6$   &    $-20.0$   &    $21.8$   &    $ 5.6$   \\ 
0841$+$1256A &  2.375		&   $ 0.6$   &    $ 9.6$   &    $ 7.4$   &    $-19.9$   &    $22.0$   &    $ 5.7$   \\ 
0841$+$1256B &  2.476		&   $ 0.9$   &    $ 9.8$   &    $ 7.5$   &    $-20.6$   &    $21.2$   &    $ 5.6$   \\ 
1151$+$06800 &  1.774		&   $ 1.7$   &    $11.2$   &    $ 8.8$   &    $-23.5$   &    $17.8$   &    $ 5.4$   \\ 
1215$+$33300 &  1.999		&   $ 0.6$   &    $ 9.7$   &    $ 7.6$   &    $-19.6$   &    $22.1$   &    $ 5.6$   \\ 
1223$+$17530 &  2.466		&   $ 0.4$   &    $10.3$   &    $ 7.9$   &    $-22.3$   &    $19.4$   &    $ 5.3$   \\ 
1229$-$02100 &  0.395		&   $ 0.4$   &    $ 9.3$   &    $ 7.9$   &    $-17.0$   &    $22.8$   &    $ 5.5$   \\ 
1247$+$26700 &  1.223		&   $ 0.4$   &    $ 8.4$   &    $ 6.8$   &    $-15.7$   &    $26.4$   &    $ 6.0$   \\ 
1328$+$30700 &  0.692		&   $ 2.8$   &    $11.5$   &    $ 9.7$   &    $-23.3$   &    $17.4$   &    $ 5.5$   \\ 
1331$+$17040 &  2.800		&   $ 0.7$   &    $10.2$   &    $ 8.1$   &    $-20.8$   &    $21.5$   &    $ 7.1$   \\ 
2206$-$1990B &  2.076		&   $ 4.0$   &    $10.9$   &    $ 8.5$   &    $-23.3$   &    $18.1$   &    $ 5.2$   \\ 
2344$+$12400 &  2.538		&   $ 0.1$   &    $ 7.8$   &    $ 5.9$   &    $-15.1$   &    $27.1$   &    $ 6.2$   \\ 
2348$-$14440 &	 2.279		&   $ 1.1$   &    $ 9.9$   &    $ 7.7$   &    $-20.7$   &    $21.1$   &    $ 5.6$   \\ 
2359$-$0220B &  2.152		&   $ 0.5$   &    $ 8.9$   &    $ 7.0$   &    $-17.6$   &    $24.3$   &    $ 5.8$   \\ 
\end{tabular}
\end{table}

\pagebreak

\vspace{0.25cm}
\begin{flushleft}{\Large\bf Figure Captions}\end{flushleft}
\vspace{0.25cm}

{\bf Figure 1} The evolution of the quasar optical luminosity function
(QOLF), and the effects of obscuration. (a) A parametric fit to the
QOLF at redshift $z=2$; (b) A pure luminosity parameterisation of the
evolution of the characteristic quasar luminosity (${\mathrm M^*_B}$);
(c) The apparent magnitude of the `knee' in the QOLF, accounting for
PLE and cosmological dimming (error bar indicates approximate overall
uncertainties); (d) The foreground optical depth required to dim an
$L_B^*$ quasar below a survey apparent magnitude limit $m_{B_{lim}} =
19$ (and uncertainties).

{\bf Figure 2} Typical dust opacity predicted from the gas phase
metallicity and the neutral gas column density. The dashed lines
correspond to typical dust optical depths $\tau_d = 1, 0.1, 0.01$. The
solid box shows the region occupied by normal galaxies (with a
possible extension into the dotted box). The data points at [-1.8,
21.8] and [-1.6, 21.8] are I Zw 18 and the LMC. The remaining data
points are all known DLAs for which metallicity measurements are
available ([Fe/H] measurements have been corrected for
$<$[Zn/Fe]$>_{DLA}$ (see \S~\ref{s:confrontation})).

{\bf Figure 3} The model redshift selection function, which is a fit
to the combined redshift survey path lengths, X(z), of the existing
large, well defined DLA surveys (see text).

{\bf Figure 4} Comparison of DLA observations (points) with the
predictions of our model (contours) using Monte Carlo simulations in
Einstein-de Sitter cosmology. The dashed lines indicate the
approximate selection effects which affect the data (derived from DLA
surveys, including the metallicity measurements of Pettini and
collaborators; see text).

{\bf Figure 5} Cumulative distributions of observed and synthetic DLAs
in (a) redshift, (b) column density and (c) metallicity in the low
density universe. Identical selection effects, empirically derived
from DLA surveys (including the metallicity measurements of Pettini
and collaborators; see text), have been applied to the model and the
data.

{\bf Figure 6} Comparison of DLA observations (points) with the
parameter space density of systems predicted by our model (contours)
using Monte Carlo simulations in low density cosmology. Identical
selection effects, empirically derived from DLA surveys (including the
metallicity measurements of Pettini and collaborators; see text), have
been applied to the model and the data.

{\bf Figure 7} Comparison of DLA metallicities as a function of
redshift from observations (points, [Zn/H] and [Fe/H]) and our model
(contours), as in Figure 6 {\it but} with no selection effects
applied.

{\bf Figure 8} The star formation history of the Universe. The data
points are for UV observations, taken from Steidel 1999 and {\it not}
corrected for extinction. Our model prediction is shown as a (smoothed)
curve. The curve has {\it not} been fitted to these data.

{\bf Figure 9} The global properties of our model Universe. The
fraction of the closure mass density supplied by baryons in galaxies,
$\Omega_{b_{in}}$ (solid line); gas in galaxies, $\Omega_{gas_{in}}$
(dashed) ; in stars, $\Omega_{stars}$ (dotted); in metals,
$\Omega_{metals}$ (dash-dot), in dust, $\Omega_{dust}$
(dash-dot-dot-dot). The contribution from total mass in galaxies,
$\Omega_{tot_{in}} = 10 \Omega_{b_{in}}$. The contribution from mass not
yet formed into galaxies, $\Omega_{tot,out} = \Omega_M -
\Omega_{tot_{in}}$, where the overall mass contribution $\Omega_M = 0.3$
in our low density cosmology. The data points are the inferred
$\Omega_{gas}$ in DLAs (\cite{SIM96,RT99,SW00}).

{\bf Figure 10} Local Group dwarf galaxies as a function of distance
from the Milky Way (a) HI gas fraction (b) Stellar metallicity [Fe/H]
for dwarf galaxies with mass $<10^8 \Msun$ (open circles) and $>10^8
\Msun$ (crosses).

{\bf Figure 11} Ratios of prompt element (Promptium, Pr) to delayed
element abundances(Delaydium, De) for simulated DLA survey (points as
dots), with data for [O/Fe] (circles), [S/Fe] (`S' shapes) and [Si/Fe]
(triangles), with errors) as a function of a) absorption redshift b)
metallicity ([De/H] or [Fe/H]).

\pagebreak

\pagebreak

\pagebreak

\pagebreak

\pagebreak

\pagebreak

\begin{figure}[p]
\centering
\resizebox{\figwidth}{!}{\rotatebox{0}{\includegraphics{figure6.eps} }}
	\caption{\label{f:figure6}}
	% Comparison of DLA observations (points) with the parameter space
	% density of systems predicted by our model (contours) using Monte
	% Carlo simulations in low density cosmology. Identical selection
	% effects, empirically derived from DLA surveys (including the
	% metallicity measurements of Pettini and collaborators; see text),
	% have been applied to the model and the data.}
%    
\end{figure}

\pagebreak

\begin{figure}[p]
\centering
\resizebox{\figwidth}{!}{\rotatebox{-90}{\includegraphics{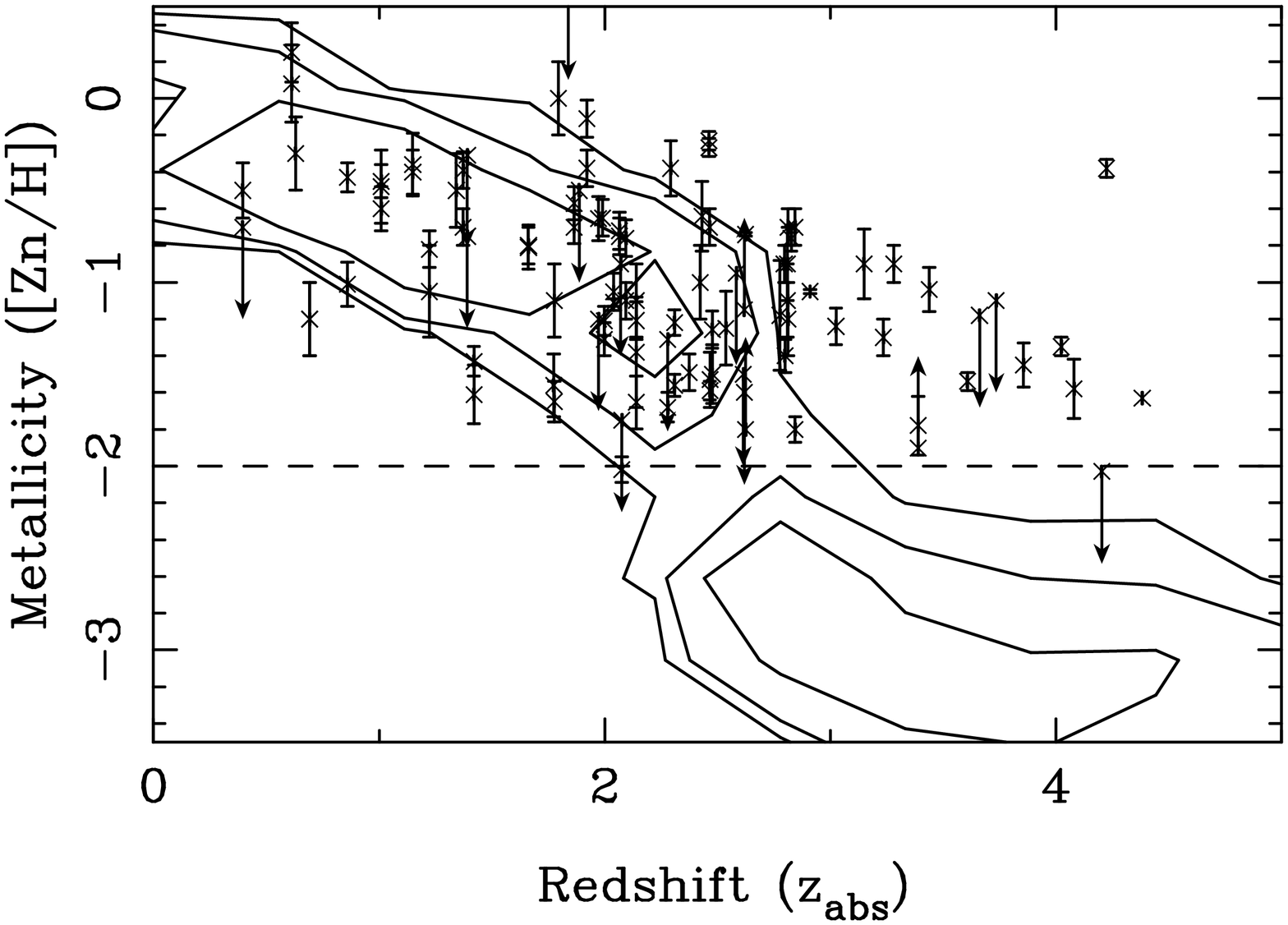} }}
	\caption{\label{f:extra_figure}}
\end{figure}

\pagebreak

\begin{figure}[p]
\centering
\resizebox{\figwidth}{!}{\rotatebox{-90}{\includegraphics{figure7.eps} }}
    \caption{\label{f:figure7}}
	% The star formation history of the Universe. The data points are
	% for UV observations, taken from Steidel 1999 and {\it not}
	% corrected for extinction. Our model prediction is shown as a
	% (smoothed) curve. The curve has {\it not} been fitted to the
	% data.}
%
\end{figure}

\pagebreak

\begin{figure}[p]
\centering
\resizebox{\figwidth}{!}{\rotatebox{-90}{\includegraphics{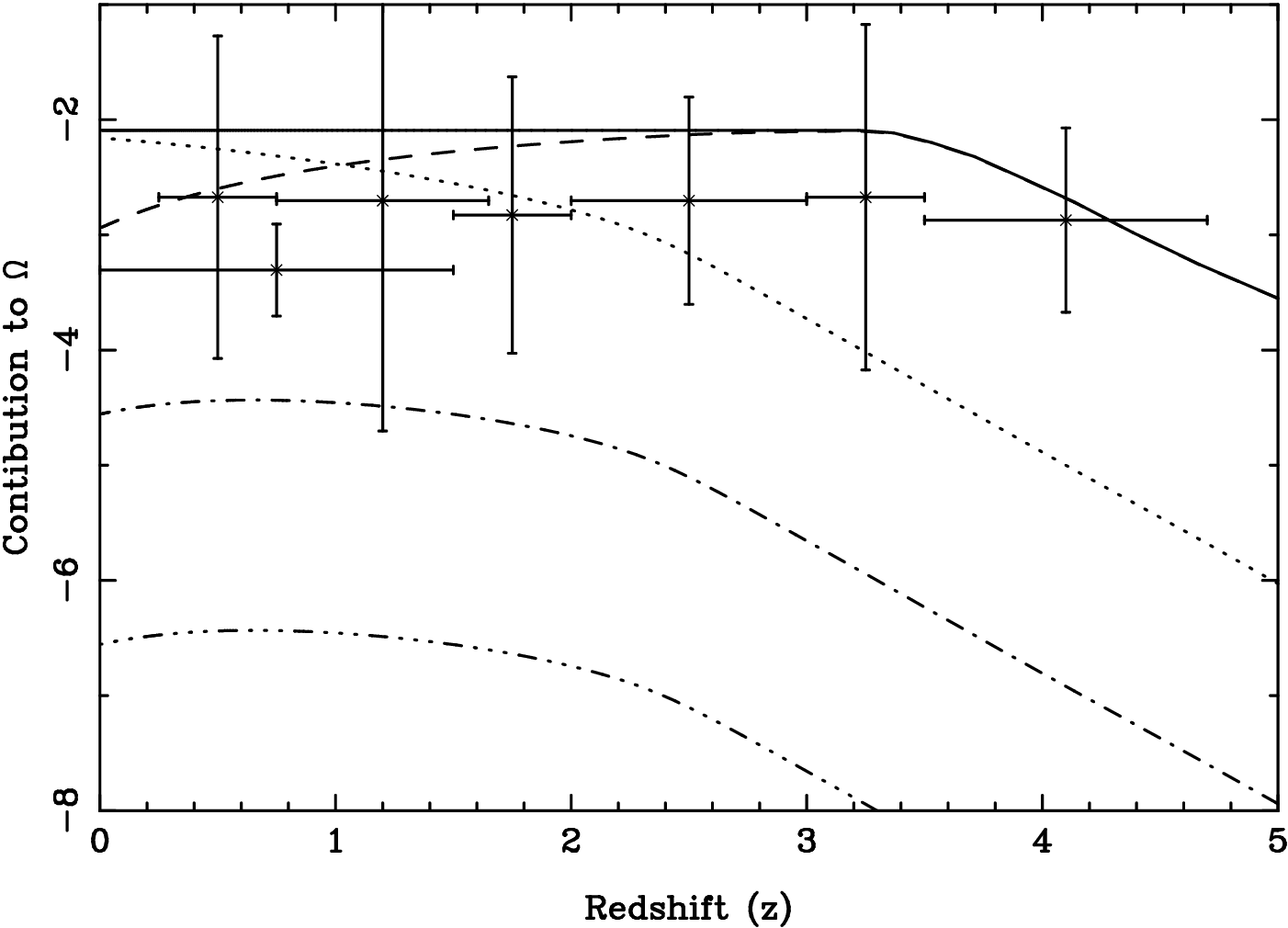}}}
	\caption{\label{f:figure8}} 
	% The global properties of our model Universe. The fraction of
	% the closure mass density supplied by baryons in galaxies,
	% $\Omega_{b,in}$ (solid line); gas in galaxies, $\Omega_{gas,in}$
	% (dashed) ; in stars, $\Omega_{stars}$ (dotted); in metals,
	% $\Omega_{metals}$ (dash-dot), in dust, $\Omega_{dust}$
	% (dash-dot-dot-dot). The contribution from total mass in galaxies,
	% $\Omega_{tot,in} = 10 \Omega_{b,in}$. The contribution from mass
	% not yet formed into galaxies, $\Omega_{tot,out} = \Omega_M -
	% \Omega_{tot,in}$, where the overall mass contribution $\Omega_M =
	% 0.3$ in our low density cosmology. The data points are the
	% inferred $\Omega_{gas}$ for DLAs (Storrie-Lombardi, McMahon and
	% Irwin 1996).}
%
\end{figure}

\pagebreak

\begin{figure}[p]
\centering
\resizebox{\figwidth}{!}{\rotatebox{0}{\includegraphics{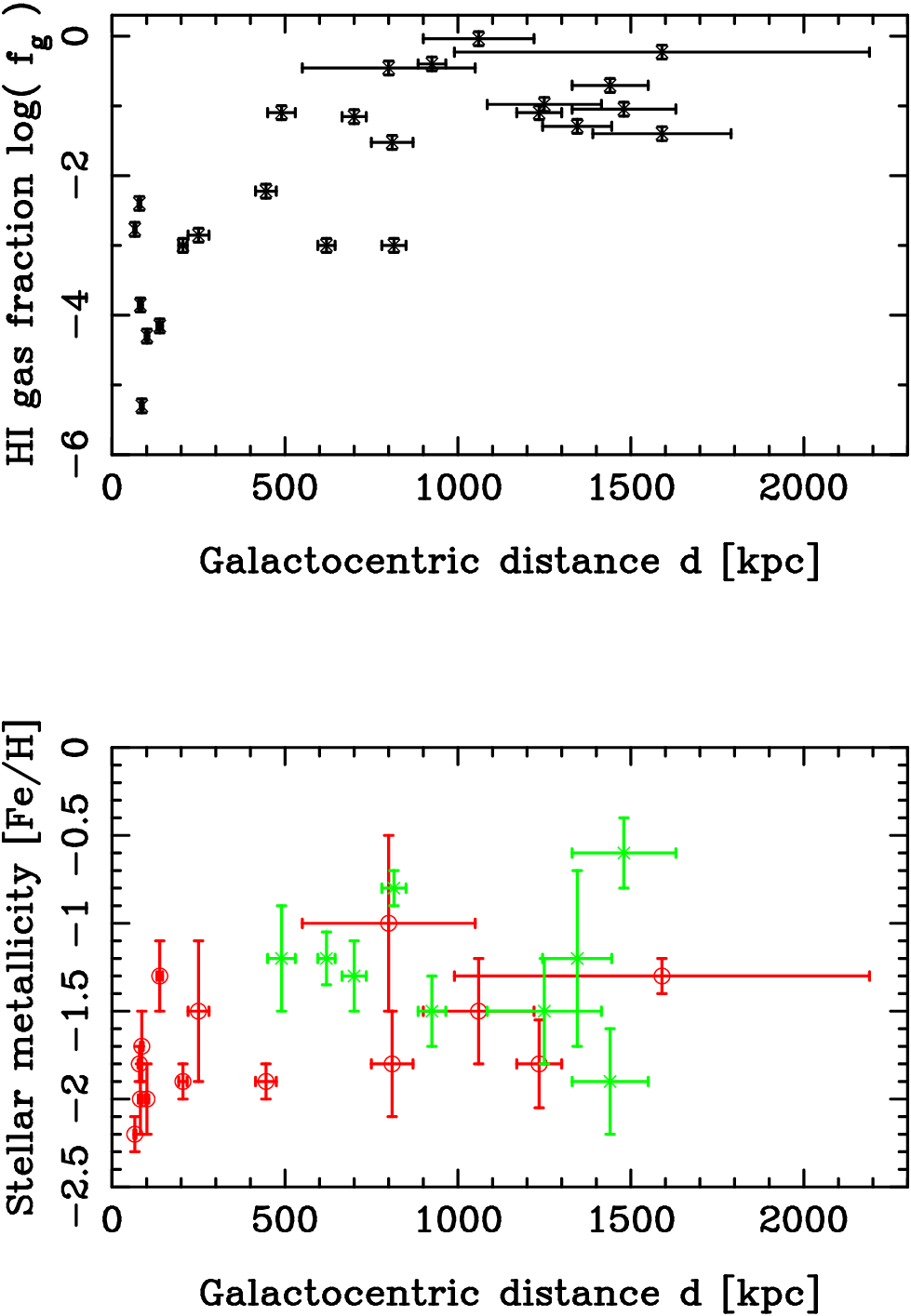} }}
	\caption{\label{f:figure9}} 
	% Local Group dwarf galaxies as a function of distance from the
	% Milky Way (a) HI gas fraction (b) Stellar metallicity [Fe/H] for
	% dwarf galaxies with mass $<10^8 M/\Msun$ (open circles) and
	% $>10^8 M/\Msun$ (crosses). }
%
\end{figure}

\pagebreak

\begin{figure}[p]
\centering
\resizebox{\figwidth}{!}{\rotatebox{0}{\includegraphics{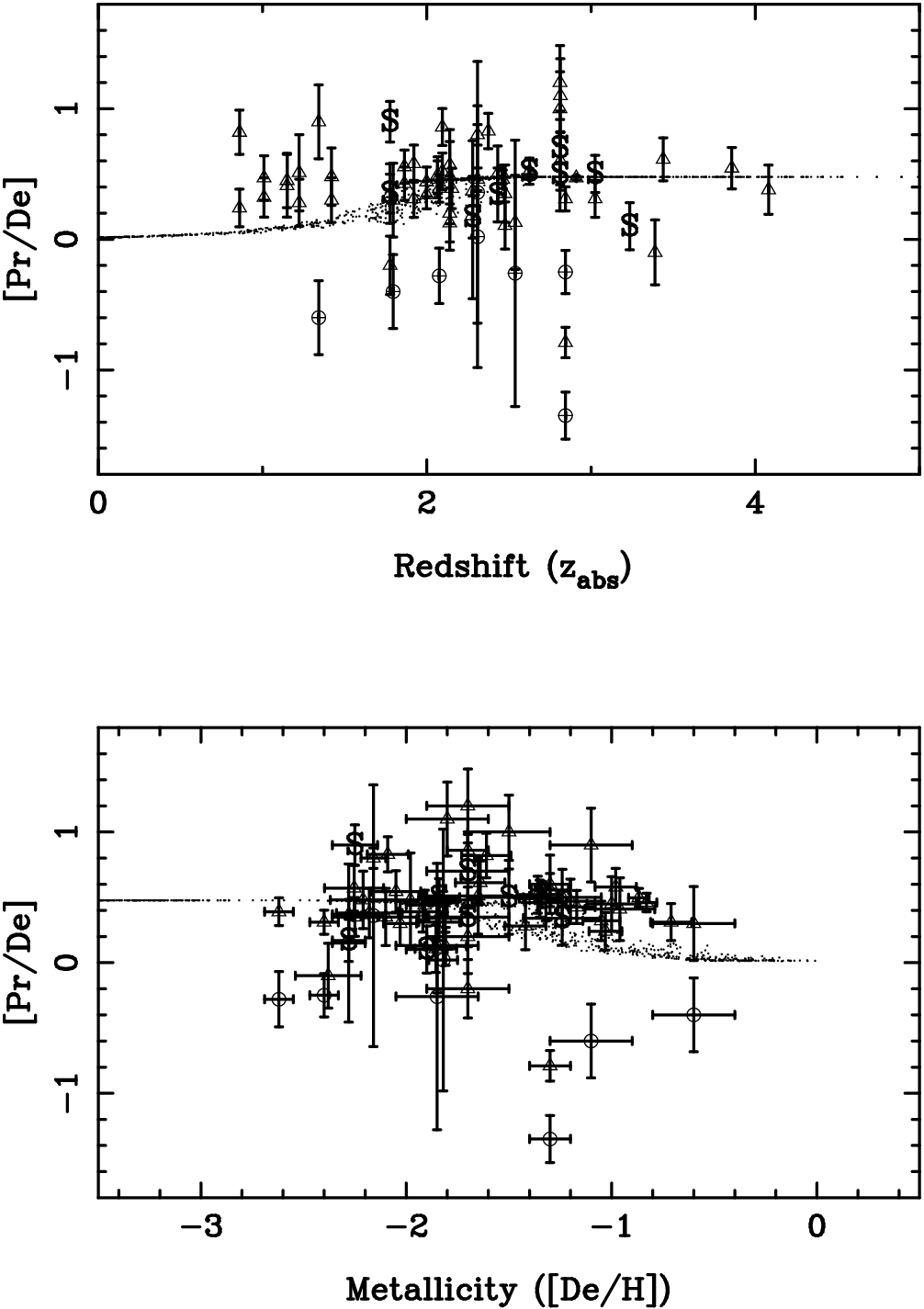} }}
    \caption{\label{f:figure10}}
    % Ratios of prompt element to delayed element abundances (simulated
    % DLA survey points as dots, with data for [O/Fe] (circles), [S/Fe]
    % (`S' shapes) and [Si/Fe] (triangles), with errors) as a function of a)
    % absorption redshift b) metallicity ([De/H] or [Fe/H]).}
%
\end{figure}

\end{spacing}

\end{document}